\documentclass[final,5p,times,twocolumn]{elsarticle}

\usepackage{amssymb}
\usepackage{amsthm}
\usepackage{amsmath}
\usepackage{array}
\usepackage{graphicx}
\usepackage{theorem}
\usepackage{dcolumn}
\usepackage{bm}
\usepackage{setspace}
\usepackage{float}
\usepackage{subcaption}

\graphicspath{{figures/}}

\newsavebox\mybox

\newcommand{\TxU}[1]{\textsuperscript{#1}}

\journal{Int. J. Heat Mass Transfer}

\usepackage{xtab}
\xentrystretch{-0.1}

\usepackage{lineno}
\begin{document}
	
\begin{frontmatter}
		
\title{Heat and mass transfer during a sudden loss of vacuum in a liquid helium cooled tube - Part III}

\author[label1,label2]{Nathaniel Garceau}
\author[label1,label2]{Shiran Bao}
\author[label1,label2]{Wei Guo\corref{c1}}
\address[label1]{National High Magnetic Field Laboratory, 1800 East Paul Dirac Drive. Tallahassee, FL 32310, USA}
\address[label2]{Mechanical Engineering Department, FAMU-FSU College of Engineering, Florida State University, Tallahassee, FL 32310, USA}
\cortext[c1]{Corresponding: wguo@magnet.fsu.edu}
		
\begin{abstract}
A sudden loss of vacuum can be catastrophic for particle accelerators. In such an event, air leaks into the liquid-helium-cooled accelerator beamline tube and condenses on its inner surface, causing rapid boiling of the helium and dangerous pressure build-up. Understanding the coupled heat and mass transfer processes is important for the design of beamline cryogenic systems. Our past experimental study on nitrogen gas propagating in a copper tube cooled by normal liquid helium (He I) has revealed a nearly exponential slowing down of the gas front. A theoretical model that accounts for the interplay of the gas dynamics and the condensation was developed, which successfully reproduced various key observations. However, since many accelerator beamlines are actually cooled by the superfluid phase of helium (He II) in which the heat transfer is via a non-classical thermal-counterflow mode, we need to extend our work to the He II cooled tube. This paper reports our systematic measurements using He II and the numerical simulations based on a modified model that accounts for the He II heat-transfer characteristics. By tuning the He II peak heat-flux parameter in our model, we have reproduced the observed gas dynamics in all experimental runs. The fine-tuned model is then utilized to reliably evaluate the heat deposition in He II. This work not only advances our understanding of condensing gas dynamics but also has practical implications to the design codes for beamline safety.
\end{abstract}
		
		
\end{frontmatter}

\section{Introduction}
Many modern particle accelerators utilize superconducting radio-frequency (SRF) cavities cooled by liquid helium (LHe) to accelerate particles~\cite{Lebrun-2014-CERN_cooling}. These SRF cavities are housed inside interconnected cryomodules and form a long LHe-cooled vacuum tube, i.e., the beamline tube~\cite{Pagani-2005-PIWS}. An accelerator can experience a catastrophic breakdown if the cavities accidently lose their vacuum to the surrounding atmosphere. For example, an accidental rupture at a cryomodule interconnect will provide the room-temperature air an open path to enter the cavity space. The air can then condense on the cavity inner surface and rapidly deposit heat to the LHe, causing violent boiling in LHe and dangerous pressure build-up in the cryomodule. Safety concerns and possible damage as a result of sudden vacuum loss have led to multiple failure studies at accelerator labs~\cite{Wiseman-1994-ACE,Seidel-2002-DESYTechRep,Ady-2014-IPACP,Ady-2014-CERNTechRep,Boeckmann-2008-PICEF22,Dalesandro-2012-AIP}. These preliminary studies revealed that the gas propagation in a freezing vacuum tube can slow down substantially as compared to that in a room-temperature tube.

In order to better understand the complex coupled heat and mass transfer processes involved in a beamline vacuum break event, pioneering work has been carried out in our cryogenics lab by Dhuley and Van Sciver via venting room-temperature nitrogen (N$_2$) gas from a buffer tank to a LHe cooled vacuum tube~\cite{Dhuley-2016-IHMT-1,Dhuley-2016-IHMT-2}. Their experiments with normal liquid helium (He I) showed that the gas-front propagation slowed down nearly exponentially. This deceleration was attributed to gas condensation to the tube wall~\cite{Dhuley-2016-IHMT-2}, but a quantitative analysis on how the condensation leads to the observed exponential slowing down was not provided. Another limitation in these early experiments is that the tube system was not suitable for measurements with the superfluid phase of liquid helium (He II). This is because after the helium bath is pumped to the He II phase, only a small section of the tube can remain immersed in He II. Furthermore, the tube section above the liquid level can be effectively cooled by the flowing helium vapor to sufficiently low temperatures such that the nitrogen gas may condense at an unknown location in this tube section. This uncertainty makes subsequent data analysis and result interpretation difficult.

In our later systematic work, a longer helical tube system was fabricated~\cite{Garceau-2017-IOP}. The tube section above the liquid surface is protected by a vacuum jacket and its temperature is always maintained to above 77 K using a feedback-controlled heater system~\cite{Garceau-2019-Cryo}. This design allows us to accurately control the starting location of the gas condensation, which is crucial for reliable measurements and for convenient comparison with model simulations. Our systematic measurements with He I confirmed the observed slowing down of the gas-front propagation~\cite{Garceau-2019-IHMT-pt1}. A theoretical model that accounts for the gas dynamics, surface condensation, and heat transfer to He I was also developed, which successfully reproduced various key observations~\cite{Bao-2020-IJHMT-pt2}. On the other hand, since many accelerator beamlines are actually cooled by He II whose heat transfer characteristics are controlled by a non-classical thermal-counterflow mechanism~\cite{Van_Sciver-2012-HeCryo}, the insights gained in our He I studies are not all directly applicable. For this reason, we now extend our work to the He II cooled tube.

This paper reports our systematic experimental measurements with He II as well as our numerical simulations of the relevant heat and mass transfer processes. In Sec.~\ref{sec:Exp}, we discuss our experimental setup, the measurement procedures, and the main observations. In Sec.~\ref{sec:model}, we present our theoretical model which consists of the conservation equations of the propagating gas, a refined model of the gas condensation, and a correlation model that reasonably describes the heat transfer in He II. The validation of the model is discussed in Sec.~\ref{sec:valid}. We show that by tuning the He II peak heat-flux parameter in our model, the observed gas motion under various inlet gas pressures and mass flow rates can be well reproduced. In Sec.~\ref{sec:Heat}, we present the calculated heat-deposition rate in He II using the fine-tuned model. This information is critical in the design codes for beamline safety. Finally, a brief summary is included in Sec.~\ref{sec:Sum}. Our systematic work provides an unprecedented characterization of the heat load accompanying the vacuum break in a He II cooled tube, which could have important implications to the design and safe operation of beamline cryogenic systems.

\section{\label{sec:Exp} Experimental Measurements}
\begin{figure}[!tb]
\centering		
\includegraphics[width=1\columnwidth]{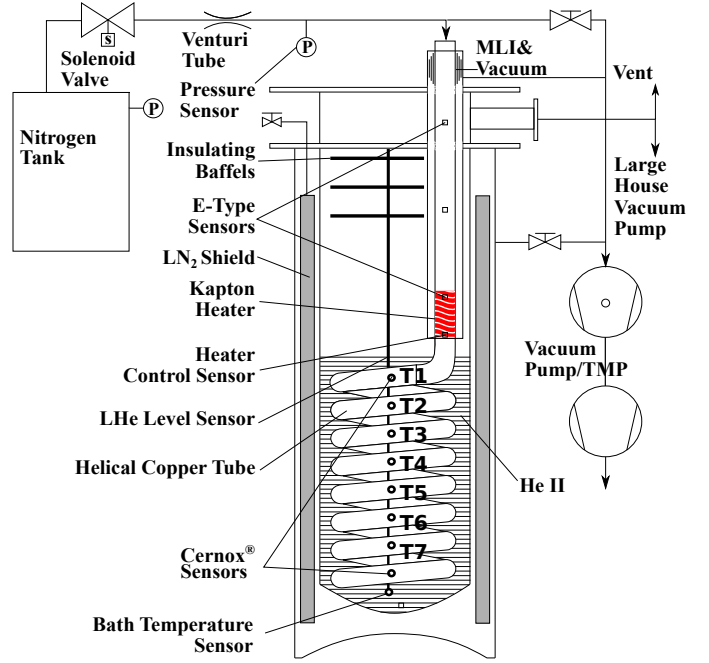}
\caption{Schematic diagram of the experimental setup used in our study on vacuum break in He~II cooled tube.\label{Fig:1}}
\end{figure}

The apparatus used in our He II experiment is the same as that in our prior He I experiment. The details of the setup can be found in~\cite{Garceau-2019-IHMT-pt1,Bao-2020-IJHMT-pt2,Garceau-2020-IOP}. In summary, the system includes a copper tube with a total length of 5.75 m. The tube has an outer diameter of $2.8$ cm and a wall thickness of 1.25~mm. It is bent into a helical coil shape with a coil diameter of 22.9 cm, as shown in Fig.~\ref{Fig:1}. The copper tube is soldered to a stainless steel extension tube which connects to the outside plumbing. The extension tube is vacuum insulated to prevent variations in temperature that could be resulted from convective cooling due to the flowing vapor during the bath pumping~\cite{Garceau-2019-Cryo}. The liquid level in the bath is controlled such that upon the completion of the pumping to a desired He II temperature, the helical copper tube is completely immersed in He II.
			
\subsection{Experimental procedures}
In our experiments, pure nitrogen gas contained in a 230-$L$ reservoir tank was used to maintain consistent flow conditions. Vacuum break was simulated by opening a fast-action solenoid valve (25 ms opening time). Following the solenoid valve, a venturi tube choked the gas flow into the evacuated tube system such that the gas velocity at the venturi exit reached the local sound speed. The gas pressures in the nitrogen tank and after the venturi were monitored using Kulite\TxU{\textregistered} XCQ-092 high speed pressure sensors. Data acquisition was conducted at a rate of 4800~Hz using LabVIEW\TxU{\textregistered} and four data acquisition modules (i.e., DT9824 from Data Translation Inc.). Measurements at four different tank pressures, i.e., 50 kPa, 100 kPa, 150 kPa and 200 kPa, were conducted.

The gas propagation was measured by monitoring the surface temperature of the copper tube using seven Lake Shore Cernox\TxU{\textregistered} temperature sensors encapsulated in 2850 FT Stycast\TxU{\textregistered} epoxy~\cite{Dhuley-2016-Cryo-cernox}. These sensors were mounted on the outer surface of copper tube using hose clamps and were placed at 72~cm apart along the tube. To reduce the thermal contact resistance between the sensors and the copper tube, indium foil and Apliezon\TxU{\textregistered}~N thermal grease were applied at the interface. To monitor the bath temperature, one Cernox sensor was allowed to float in the center of the helium bath, as shown in Fig.~\ref{Fig:1}.

\subsection{Main observations}
\begin{figure}[!tb]
\centering		
\includegraphics[width=0.9\columnwidth]{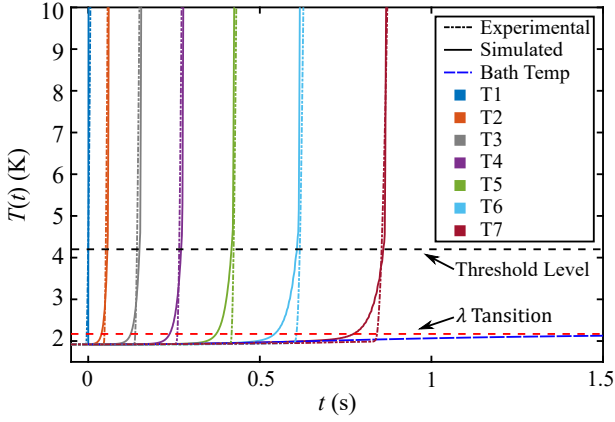}	
\caption{Measured and simulated temperature curves at different sensor locations for the experimental run at 150~kPa tank pressure.\label{Fig:2}}	
\end{figure}

After the solenoid valve was opened, the nitrogen gas flushed into the evacuated copper tube and condensed on the tube inner surface, causing variations of the tube-wall temperature. Fig.~\ref{Fig:2} shows the temperature curves recorded by the Cernox sensors in a representative experimental run with a tank pressure of 150~kPa. These temperature curves initially remain at the bath temperature (i.e., 1.95 K). Upon the arrival of the gas propagating font, the temperature curves rise abruptly to a saturation temperature of about 50~K to 60~K (not shown in the figure). Due to the heat deposited in the bath, the bath temperature $T_{bath}$ also rises steadily. Nevertheless, according to the $T_{bath}$ curve, the LHe remained in the He II phase since its temperature was below the lambda-point temperature (i.e., $T_{\lambda}=2.17$ K) over the course of the experiment. Therefore, the He II heat transfer characteristics are needed in our result analysis.

\begin{figure}[htb]	
\centering	
\includegraphics[width=0.9\columnwidth]{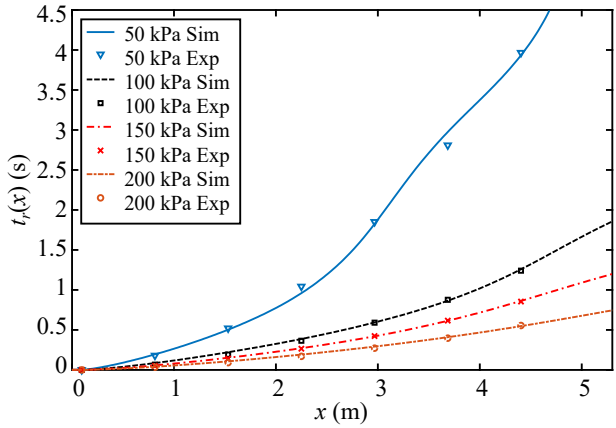}	
\caption{Measured and simulated rise times of the temperature curves at different sensor locations for experimental runs with different tank pressures. \label{Fig:3}}
\end{figure}

In order to extract quantitative information about the gas propagation, we introduce the rise time $t_r(x)$ defined as the moment when the temperature at a location $x$ from the tube entrance rises to above a threshold level of 4.2~K, as indicated in Fig.~\ref{Fig:2}. The rise time recorded at different sensor locations for all the four runs with tank pressures of 50~kPa, 100~kPa, 150~kPa, and 200~kPa are collected in Fig.~\ref{Fig:3}. At 50~kPa tank pressure and hence the lowest tube inlet mass flow rate, the slowest propagation and the strongest deceleration of the gas flow were observed. On the other hand, at 200~kPa tank pressure, the mean velocity of the gas is the highest (i.e., about 10~m/s) and the slowing-down effect is relatively mild.

\section{\label{sec:model} Theoretical model}
We describe the N$_2$ gas dynamics and the heat transfer in our He II cooled tube using an one-dimensional (1D) model similar to that discussed in our previous work~\cite{Bao-2020-IJHMT-pt2}. This 1D modeling is reasonable considering the small diameter-to-length ratio of the tube. A refined gas condensation model and a heat-transfer correlation model applicable to He II are incorporated in the current work. In what follows, we outline the relevant details.

\subsection{Gas dynamics}
The propagation of the N$_2$ gas in the tube and the evolution of its properties can be described by the conservation equations of the N$_2$ mass, momentum, and energy, as listed below:
\begin{equation}
\frac{\partial\rho_g}{\partial t}+\frac{\partial}{\partial x}(\rho_g v) = -\frac{4}{D_1} \dot{m}_{c},
\label{eq:modConsMass}
\end{equation}
\begin{equation}
				\frac{\partial}{\partial t}(\rho_g v) + \frac{\partial}{\partial x}(\rho_g v^2) = -\frac{\partial 	P}{\partial x} - \frac{4}{D_1}\dot{m}_{c} v,
				\label{eq:modConsMome}
\end{equation}
\begin{equation}
\begin{aligned}
\frac{\partial}{\partial t}\left[\rho_g\left(\varepsilon_g+ \frac{1}{2} v^2\right)\right]+\frac{\partial}{\partial x}\left[\rho_g v \left(\varepsilon_g+\frac{1}{2} v^2 + \frac{P}{\rho_g}\right)\right]=\\
-\frac{4}{D_1} \dot{m}_{c} \left(\varepsilon_g+\frac{1}{2}v^2 + \frac{P}{\rho_g} \right) - \frac{4}{D^2_1} Nu\cdot k_g (T_g-T_s).
\label{eq:modConsEner}
\end{aligned}
\end{equation}
The definitions of the parameters involved in the above equations are provided in the Nomenclature table. These equations, including the Nusselt number $Nu$ correlation for gas convective heat transfer, are the same as in our prior modeling work~\cite{Bao-2020-IJHMT-pt2}. The model also assumes that the N$_2$ gas obeys the ideal-gas equation of state:
\begin{equation}
P M_g =\rho_g R T_g,
\label{eq:IG}
\end{equation}
which is justified since the compressibility of the N$_2$ gas is close to unity in the entire experiment~\cite{Bao-2020-IJHMT-pt2}. The terms containing $\dot{m}_c$ on the right hand side of the Eq.~(\ref{eq:modConsMass})-(\ref{eq:modConsEner}) account for the effects due to the condensation of the $N_2$ gas on the tube inner surface. The parameter $\dot{m}_c$ denotes the mass deposition rate per unit tube inner surface area, which requires an additional equation to describe its time variation.

\subsection{Mass deposition}
In our previous work~\cite{Bao-2020-IJHMT-pt2}, $\dot{m}_c$ is evaluated using a sticking coefficient model derived assuming ideal Maxwellian velocity distribution of the gas molecules. Although this model is designed for use in the free molecular flow regime, it turns out to describe well the mass deposition in our He I experiments where the N$_2$ gas was essentially in the continuum flow regime~\cite{Garceau-2019-IHMT-pt1}. Nevertheless, potential inaccuracy may still occur when $\dot{m}_c$ is so large that the distribution of the molecule velocity starts to deviate from the Maxwellian distribution due to the mean flow towards the wall as caused by condensation.

To account for this effect, in the current work we adopt a refined model using the Hertz-Knudsen relation with the Schrage modification~\cite{Collier-1994-ConvBoilCond}. This kinetic theory model evaluates $\dot{m}_c$ based on the difference between the flux of the molecules condensing on the cold surface and the flux of the molecules getting evaporated from the frost layer:
\begin{equation}
\dot{m}_{c} = \sqrt{ \frac{M_g}{2\pi R}}\left({\Gamma\sigma_c}\frac{P}{\sqrt{T_g}}-\sigma_e\frac{P_w}{\sqrt{T_w}} \right),
\label{eq:mc}
\end{equation}
where \(P_w\) is the saturated vapor pressure at the wall temperature $T_w$. $\sigma_c$ and $\sigma_e$ are empirical condensation and evaporation coefficients which are typically about the same and close to unity for a very cold surface~\cite{persad-2016_Chem.Rev.}. We take 0.95 for both coefficients in our later numerical simulations. The coefficient $\Gamma$ is responsible for any deviation from the Maxwellian velocity distribution of the molecules as caused by the condensation on the wall. In terms of the mean-flow velocity towards to the cold wall $u=\dot{m}_{c}/\rho_g$, $\Gamma$ can be calculated as~\cite{Collier-1994-ConvBoilCond}:
\begin{equation}
\Gamma(\beta) = exp\left({-\beta^2}\right) +\beta\sqrt{\pi} \left[1+erf\left( {\beta}\right)\right],
\end{equation}
where $\beta=u/u_T$ with $u_T=\sqrt{2RT_g/M_g}$ being the thermal velocity of the gas molecules. Note that since $\Gamma$ depends on $\dot{m}_c$ through $u$, the Eq.~\ref{eq:mc} needs to be solved self-consistently at every time step to obtain the evolution of $\dot{m}_c$.

\subsection{Radial heat transfer}
In the governing equations for the gas dynamics and $\dot{m}_c$, the wall temperature $T_w$ and the surface temperature $T_s$ of the frost layer are needed. This information be can obtained through the analysis of the heat transfer in the radial direction.

First, assuming that the frost-layer thickness $\delta$ is small such that a linear temperature profile exists in the layer, we can evaluated the frost-layer center temperature $T_\text{c}=(T_\text{w}+T_\text{s})/2$ as~\cite{Bao-2020-IJHMT-pt2}:
\begin{equation}
\rho_{SN} C_{SN} \delta \frac{\partial T_c}{\partial t} = q_\text{dep} - q_i
\label{eq:frost}
\end{equation}
where $q_\text{dep}=\dot{m}_c[v^2/2 + {\hat{h}}_\text{g} - {\hat{h}}_\text{s}] + Nu\cdot k_g(T_\text{g}-T_\text{s})/D_1$ is the total incoming heat flux to the frost layer, and $q_i=k_\text{SN}(T_\text{s}-T_\text{w})/\delta$ denotes the outgoing heat flux to the copper tube wall. The growth rate of the frost layer thickness $\delta$ is given by ${\dot{\delta}=\dot{m}_c/\rho_\text{SN}}$.

Then, the variation of the copper tube wall temperature $T_w$ can be described by:
\begin{equation}
\rho_w C_{w} \frac{D^2_2 - D^2_1}{4D_1} \frac{\partial T_w}{\partial t} = q_i - q_\text{He} 	\frac{D_2}{D_1}+\frac{D^2_2 - D^2_1}{4D_1} k_w \frac{\partial^2 T_w}{\partial x^2}.
\label{eq:wall}
\end{equation}
In this equation, we neglected the temperature gradient across the thickness of the copper tube, which is justified considering the high thermal conductivity of copper and the small wall thickness. The parameter $q_\text{He}$ denotes the heat flux from the copper tube to the He II bath, which is modeled separately.

\subsection{He II heat transfer modeling}
It has been known that He II can be treated as a mixture of two miscible fluid components: an inviscid and zero-entropy superfluid and a viscous normal fluid~\cite{Tilley-book}. Heat transfer in bulk He II is via an extremely effective counterflow mode instead of classical convection~\cite{Van_Sciver-2012-HeCryo}: the normal fluid carries the heat away from the heat source at a velocity proportional to the heat flux, whereas the superfluid moves in the opposite direction to ensure mass conservation. When the heat flux exceeds a small critical value, quantized vortices are nucleated spontaneously in the superfluid, which can impede the counterflow and increase the temperature gradient~\cite{Vinen-1957-PRS-III}. Our lab has done extensive characterization work on both steady-state~\cite{Marakov-2015-PRB,Gao-2016-JETP,Gao-2017-PRB,Gao-2017-JLTP,Bao-2018-PRB,Mastracci-2018-PRF,Mastracci-2019-PRF,Mastracci-2019-PRF-II} and transient counterflow~\cite{Bao-2019_PRApplied,Bao-2020-IJHMT,Bao-2021-PRB} in bulk He II.

\begin{figure}[!tb]
\centering
\includegraphics[width=0.9\columnwidth]{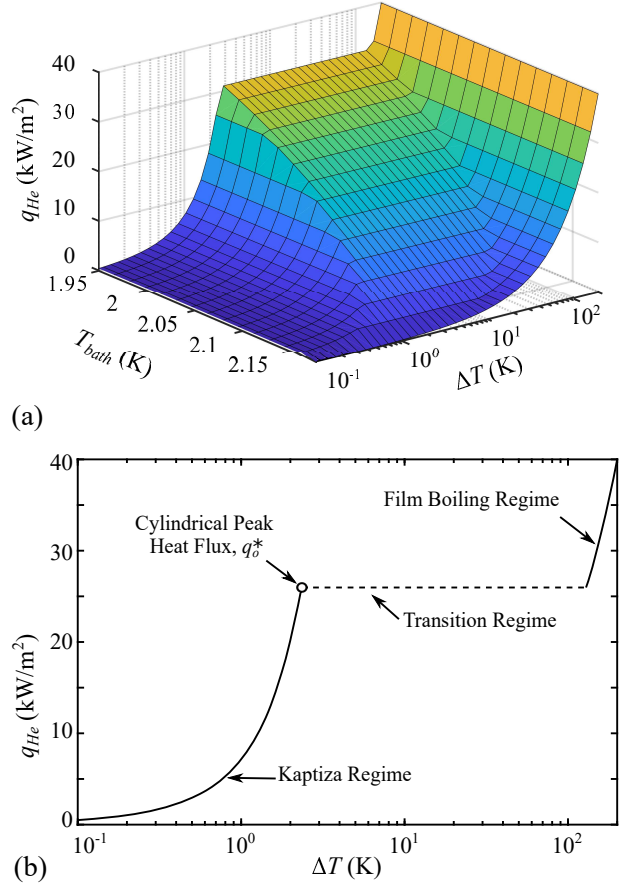}\\
\caption{(a) The heat flux $q_\text{He}$ from the copper tube to the He II bath as a function of $T_{bath}$ and the temperature difference $\Delta T=T_w-T_{bath}$, calculated using the model outlined in the text with $m=3.2$. (b) A sliced view of (a) at $T_{bath}=2$~K.\label{Fig:4}}
\end{figure}

Nonetheless, when the heat transfer from a solid surface to the He II bath is concerned, the situation becomes complicated. There are three distinct regimes of heat transfer, i.e., the Kaptiza regime, the transition regime, and the film boiling regime~\cite{Van_Sciver-2012-HeCryo}. The Kaptiza regime occurs at low heat fluxes where the copper tube wall is in full contact with the He II. In this regime, the heat flux $q_\text{He}$ depends on the temperatures $T_w$ and $T_{bath}$ through the following empirical correlation~\cite{Kashani-1985-Cryo}:
\begin{equation}
q_\text{He}=\alpha\left(T^{n}_{w}-T^{n}_{bath}\right),
\end{equation}
where the parameters $\alpha$ and $n$ are normally determined experimentally. For copper oxidized in air, measurements gave $\alpha~=~2.68$~kW/m$^2$K$^n$ and $n=2.46$~\cite{Kashani-1985-Cryo}.
			
As $q_\text{He}$ increases to the so-called peak heat flux $q^{*}_{0}$, the temperature gradient associated with the counterflow in He II builds up such that the He II temperature adjacent to the tube surface can reach the lambda point $T_{\lambda}$. Consequently, vapor bubbles start to nucleate on the tube surface, and the transition regime starts. In this regime, $q_\text{He}$ is essentially limited to $q^{*}_{0}$ while the temperature difference $\Delta T=T_w-T_{bath}$ increases. For a steady counterflow from a cylindrical surface with a diameter $D_2$, $q^{*}_{0}$ depends on $T_{bath}$ as~\cite{Van_Sciver-2012-HeCryo}:
\begin{equation}
q^{*}_{0}(T_{bath})=\left( \frac{m-1}{D_2/2}\int_{T_{bath}}^{T_{\lambda}}\frac{dT}{f(T)} \right)^{1/m}
\label{Eq:peakHeat}
\end{equation}	
where the He II heat conduction function $f(T)^{-1}$ has been tabulated~\cite{Van_Sciver-2012-HeCryo}. The Gorter-Mellink exponent $m$ has a theoretical value of 3 but has been shown to vary experimentally in the range of 3 to 4~\cite{Arp-1970-cryo,Sato-2006-ACE,Ahlers-1969-PRL,Chase-1962-PR}. We note that the above correlation is derived for steady counterflow. In a highly transient process, the instantaneous heat flux can be affected by processes such as the buildup of the vortices and the emission of thermal waves~\cite{Bao-2021-PRB}. For simplicity, we shall still adopt Eq.~\ref{Eq:peakHeat} but will treat $m$ as a tuning parameter so the model can be fine-tuned to better describe the observations.

At sufficiently large $\Delta T$, the vapor bubbles merge together and form a thin film that covers the tube surface. In this film boiling regime, $q_\text{He}$ can be estimated as~\cite{Van_Sciver-2012-HeCryo}:
\begin{equation}
q_\text{He}=h_{film}(T_w-T_{bath}),
\end{equation}
where the film boiling coefficient $h_{film}$ is expected to depend on the radius of the tube and the hydrostatic head pressure, but relevant experimental information is scarce. For cylindrical heaters with diameters over 10 mm placed at a depth of 10~cm in He~II, limited measurements suggest $h_{film}\simeq200$~W/m$^2$~\cite{Bretts-1975-ACE}. We shall adopt this $h_{film}$ value in all later simulations. Indeed, the exact value of $h_{film}$ only has a minor effect on the analysis results, since the relevant heat transfer should be mainly in the Kaptiza and the transition regimes, as judged from the observed $\Delta T$ (also see discussions in Sec.~\ref{sec:Heat}).

Based on the model that we have outlined, we can calculate $q_\text{He}$ and plot it as a function of $T_{bath}$ and $\Delta T$. A representative graph obtained with $m=3.2$ is shown in Fig~\ref{Fig:4}~(a). A sliced view of $q_\text{He}$ at $T_{bath}=2$ K is included in Fig~\ref{Fig:4}~(b), where the different heat transfer regimes are labeled for clarity.

\section{\label{sec:valid} Model validation and fine-tuning}
We have conducted numerical simulations using the governing equations as outlined in the previous section. These equations are solved using a two-step first-order Godunov-type finite-difference method~\cite{Danaila-2007-IntroSciComp,Sod-1978-JCP} in the same computational domain as in our previous work~\cite{Bao-2020-IJHMT-pt2}. Note that due to the finite volume of the He II in our experiment, the bath temperature $T_{bath}$ was observed to rise by about 0.1-0.15~K in each run due to the heat deposition. Since the parameter $q^{*}_{0}$ in our model depends strongly on $T_{bath}$, we have implemented a time varying $T_{bath}$ in our simulation using the experimental $T_{bath}$ data.

\begin{figure}[!tb]
\centering		
\includegraphics[width=0.9\columnwidth]{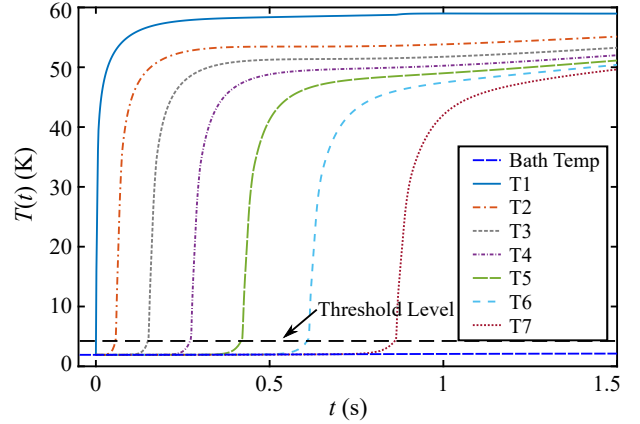}
\caption{Simulated tube-wall temperature curves at the sensor locations at a tank pressure of 150~kPa. The optimal $m=3.08$ is used in the model.}
\label{Fig:5}
\end{figure}

In Fig.~\ref{Fig:5}, we show the calculated tube-wall temperature curves at the sensor locations for a representative run with a tank pressure of 150 kPa. The Gorter-Mellink exponent $m=3.08$ was used in this calculation. These temperature curves rise sharply to about 50-60 K upon the arrival of the gas front. The separation between adjacent temperature curves gradually increases, suggesting a slowing down of the propagation. These features all agree well with the experimental observations. The calculated temperature curves are also included in Fig.~\ref{Fig:2}, where a quantitative agreement with the measured data can be clearly seen.

\begin{figure}[!tb]		
\centering
\includegraphics[width=0.9\columnwidth]{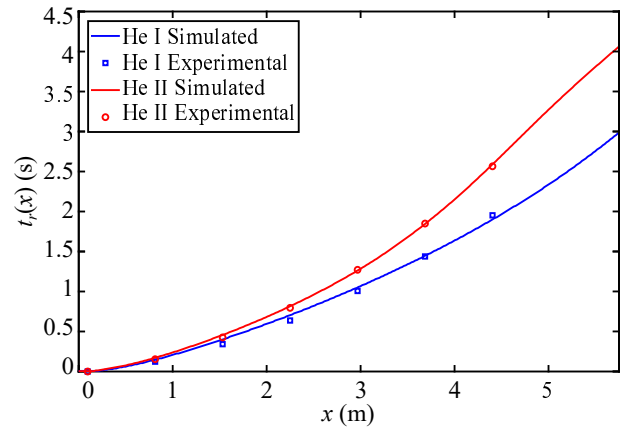}	
\caption{Comparison of rise-time data in our He~I and He~II experiments with the same tube system at a tank pressure of 150~kPa. \label{Fig:6}}
\end{figure}
{\centering	
\begin{table}[!tb]
\caption{Optimized Gorter-Mellink exponent $m$ for He II runs at different tank pressures.\label{Tab:ms}}
\centering
\begin{tabular}{c c c}	
					\hline
					Tank Pressure (kPa) 	& Optimal $m$ 	\\
					\hline
					50       	    	& 3.26    			\\
					100      	 		& 3.21    		 \\
					150           		& 3.08    		  \\
					200           		& 3.05    	     \\
					\hline
				
\end{tabular}
\end{table}
}

In Fig.~\ref{Fig:3}, we include the calculated rise-time curves to compare with the experimental data obtained at various tank pressures. The profiles of these calculated curves depend on the choice of the $m$ value. For each pressure, we tune the Gorter-Mellink exponent $m$ in the range of 3 to 4 to find the optimal value that minimizes the total variance between the experimental rise times and the simulated rise times summed over all sensor locations. The $m$ values obtained through this optimization procedure at different tank pressures are listed in Table~\ref{Tab:ms}. As one can see in Fig.~\ref{Fig:3}, the optimized rise-time curves agree nicely with the experimental data in all the runs, which thereby confirms the fidelity of the model. This fine-tubed model with the optimal $m$ exponent will be use in our later calculation of the heat deposition in He II.

We would also like to briefly comment on the difference of the gas dynamics between our He I and He II experiments. Fig.~\ref{Fig:6} shows a comparison of He~I and He~II rise-time data obtained using the same upgraded tube system at an tank pressure of 150~kPa. The simulated curves are also included. It is clear that the He~II curve shows a stronger slowing effect compared to the He~I curve. This difference was suggested in early preliminary studies~\cite{Dhuley-2016-Dis} but the result was compromised by various issues as discussed in Ref.~\cite{Garceau-2019-IHMT-pt1}. Now we can draw a reliable conclusion that the He II cooled tube indeed has a stronger slowing effect to the propagating gas due to the more effective heat transfer in the bath.

\section{\label{sec:Heat} Heat Deposition in He II}
\begin{figure}[!tb]		
\centering
\includegraphics[width=0.9\columnwidth]{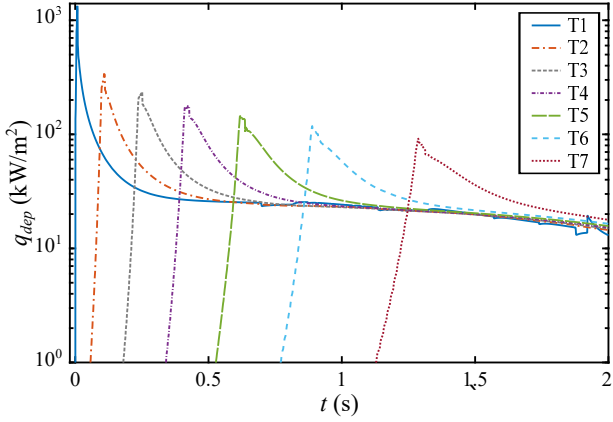}	
\caption{Simulated total heat flux $q_\text{dep}$ deposited to the tube wall at the sensor locations for a representative run with a tank pressure of 100~kPa.\label{Fig:7}}
\end{figure}

Let us first discuss the total heat flux $q_\text{dep}$ to the tube wall at the sensor locations. Fig.~\ref{Fig:7} shows the $q_\text{dep}(t)$ curves calculated using the refined model for a representative run with 100 kPa tank pressure. We see that all the $q_\text{dep}$ curves first spike to above 10$^2$ kW/m$^2$ and then collapse to a nearly universal curve that slowly drops with time to about 20 kW/m$^2$. This spiking is associated with the arrival of the gas front, since the mass deposition rate (and hence $q_\text{dep}$) is the highest when a clean and cold wall area is first exposed to the gas. As the frost layer grows, the tube wall temperature $T_w$ rises sharply to a saturation level (see Fig.~\ref{Fig:5}). Consequently, the mass deposition rate drops to a roughly constant level at all the sensor locations, which leads to the convergence of the $q_\text{dep}$ curves.
			
\begin{figure}[!tb]
\centering		
\includegraphics[width=0.9\columnwidth]{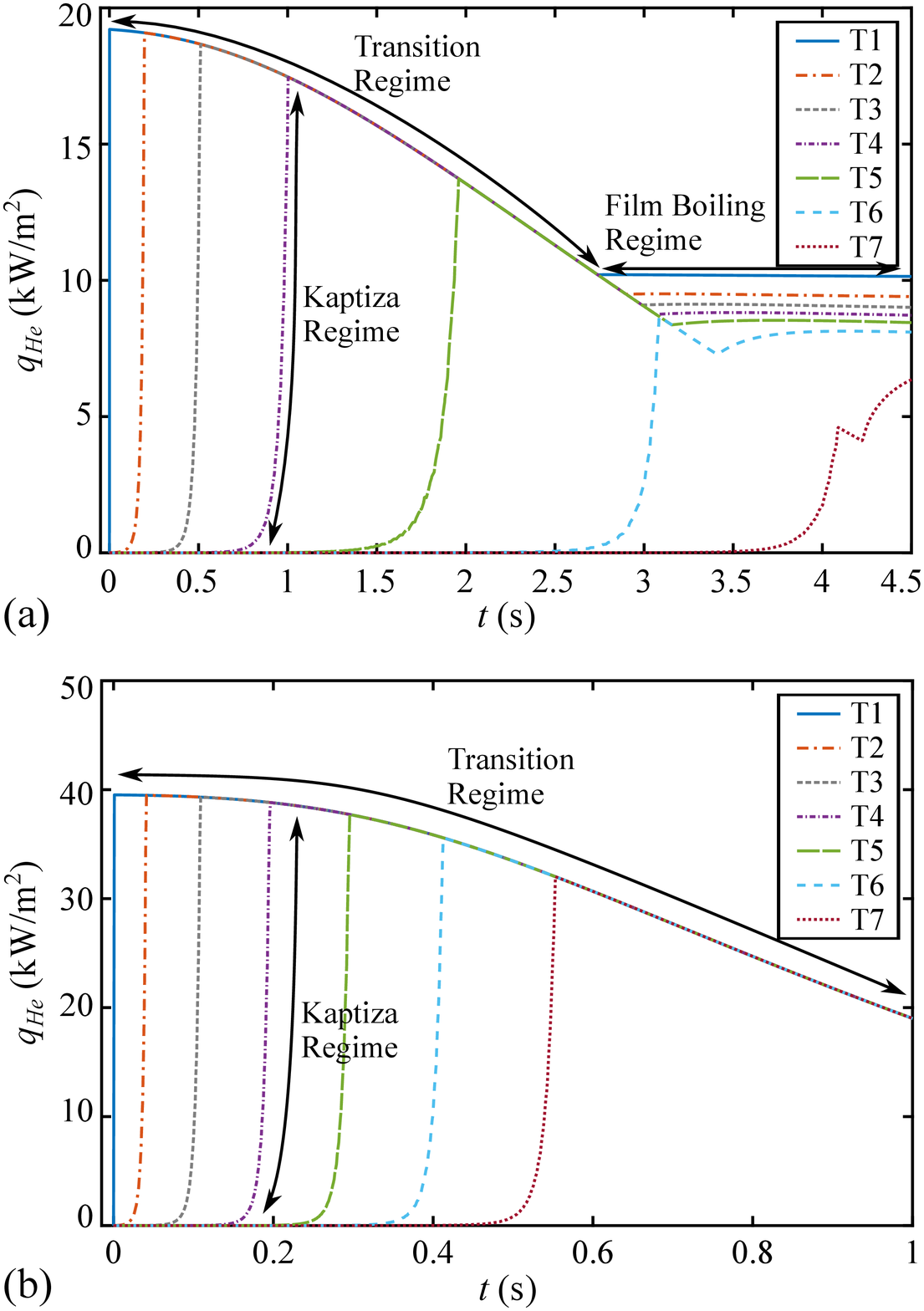}
\caption{Simulated heat flux into He II at the sensor locations for the runs with (a) 50~kPa tank pressure; and (b) 200~kPa tank pressure. \label{Fig:8}}
\end{figure}

Next, we evaluate the heat flux $q_\text{He}$ into the He II bath. Fig.~\ref{Fig:8}~(a) and (b) show $q_\text{He}$ over time at the sensor locations for the runs with the lowest tank pressure (i.e., 50~kPa) and with the highest tank pressure (i.e., 200~kPa), respectively. In both runs, $q_\text{He}$ spikes up again upon the arrival of the gas front, and the heat transfer evolves quickly from the Kapitza regime to the transition regime. In the transition regime, $q_\text{He}$ is limited to the peak heat flux $q^{*}_{0}(T_{bath})$, and therefore all the $q_\text{He}$ curves at different sensor locations merge into a universal curve following the spiking. The gradual decrease of the merged $q_\text{He}$ curves is due to the decrease in $q^{*}_{0}(T_{bath})$ as the bath temperature $T_{bath}$ slowly increases. A unique feature observed in the 50~kPa run is that after about 2.5 s, the heat flux $q_\text{He}$ saturates to a nearly constant level controlled by the film boiling. This film boiling regime is not observed in the other runs by the time the gas front reaches the last sensor.

\begin{figure}[!tb]		
\centering
\includegraphics[width=0.9\columnwidth]{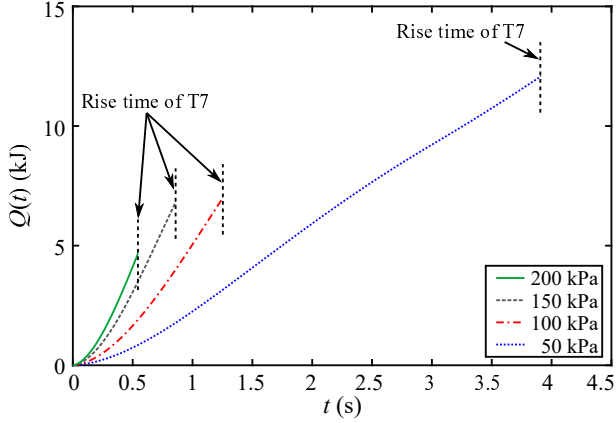}	
\caption{Calculated total heat $Q(t)$ deposited in the He II bath under different tank pressures. The vertical dashed line marks the time when the gas front reaches the last sensor. \label{Fig:9}}
\end{figure}

To understand this difference, we refer to the He II heat transfer model as shown in Fig.~\ref{Fig:4}. When the bath temperature $T_{bath}$ is low, the transition regime and the film boiling regime intersect at a $\Delta T$ value that is typically above the maximum $\Delta T$ observed over the course of the gas prorogation. But as the heat keeps being deposited in the bath, $T_{bath}$ increases and hence $q^{*}_{0}$ decreases, which lowers the intersection $\Delta T$ between the two regimes. This effect is especially pronounced for the case with the smallest $m$ exponent, i.e., the 50~kPa run. Therefore, it is most likely to observe the film boiling regime in the 50~kPa run if the heat transfer to the bath in this run is relatively large. To check this point, we have calculated the total heat deposited in the bath from the entire tube as a function of time:
\begin{equation}
Q(t)=\int^t_0dt'\int^L_0dx'\pi D_2\cdot q_\text{He}(x',t').
\end{equation}
The results for all the four runs are shown in Fig.~\ref{Fig:9}. It is clear that despite the the smallest heat deposition rate in the 50~kPa run, the highest amount of heat (i.e., over 12 kJ) is transferred to the bath, which is simply the consequence of the longest heat-deposition time before the gas front reaches the last sensor.

\section{\label{sec:Sum} Summary}
We have conducted systematic experimental and numerical studies of N$_2$ gas propagation and condensation inside an evacuated copper tube cooled by He II. By adjusting the Gorter-Mellink exponent $m$ in our theoretical model, we have nicely reproduced the observed tube-wall temperature variations in all the runs at different tank pressures. The fine-tuned model is then used to calculate the heat flux to the He II bath. Our results reveal that upon the arrival of the gas front, the heat flux spikes up to the peak heat flux at which nucleate boiling occurs on the tube outer surface. This heat flux appears to be higher at larger tank pressure (and hence larger inlet mass flow rate and mass deposition rate). Nevertheless, before the gas front reaches the tube end, the highest amount total heat is deposited in the He II bath in the run with the lowest tank pressure. The knowledge we have produced in this study may benefit the design of future beamline systems cooled by He II.

We would also like to point out that in our study we have observed a more or less constant mass deposition rate in the tube section where the gas front has passed. Similar phenomenon was observed in our previous He I work as well~\cite{Bao-2020-IJHMT-pt2}. These observations suggest that in a sufficiently long tube such as an accelerator beamline tube, the gas flushing into the tube would eventually be almost consumed by the mass deposition on the tube inner wall after a certain propagation range. This maximum range of frost contamination is a useful parameter in beamline research and design. We plan to explain the physical mechanism underlying this nearly constant mass deposition rate and present a simple correlation that allows reliable estimation of the maximum contamination range in a future publication.

\section*{Acknowledgments}
The authors would like to thank Prof. S. W. Van Sciver for valuable discussions. This research is supported by the U.S. Department of Energy under Grant No. DE-SC0020113 and was conducted at the National High Magnetic Field Laboratory at Florida State University, which is supported through the National Science Foundation Cooperative Agreement No. DMR-1644779 and the state of Florida.

	
\section*{Nomenclature}
\begin{xtabular}{cp{5cm}p{1.5 cm}}
\hline
\hline
Variable 		& Description											& Units				    \\
$C$       	    & Specific heat                                        	& J/(kg$\cdot $K)   	\\
$D_1$           & Inner diameter of the tube                           	& m                 	\\
$D_2$           & Outer diameter of the tube                           	& m                 	\\
$k$             & Thermal conductivity                                 	& W/(m$\cdot $K)    	\\
$\hat{h}$       & Specific enthalpy                                  	& J/kg    	            \\
$h_{film}$		& Film boiling coefficient							   	& W/(m$^2\cdot $K)  	\\
$m$       		& Gorter-Mellink exponent                              	&              	        \\
$\dot m$        & Mass flow rate                                       	& kg/s              	\\
$\dot m_c$      & Mass deposition rate                                 	& kg/(m$^2\cdot $s) 	\\
$M_g$           & Gas molar mass                                      	& kg/mol            	\\
$Nu$            & Nusselt number                                       	&                   	\\
$P$             & Pressure                                            	& Pa                	\\
$q_{dep}$       & Deposition heat flux                                 	& W/m$^2$           	\\
$q_{He}$        & Heat flux to the liquid helium bath                  	& W/m$^2$           	\\
$q_{i}$         & Heat flux to the inner tube surface                  	& W/m$^2$           	\\
$q^*_0$         & Peak heat flux                                     	& W/m$^2$           	\\
$R$             & Ideal gas constant                                   	& J/(mol$\cdot $K)  	\\
$t$             & Time                                                 	& s                 	\\
$t_r$           & Rise time                                            	& s                 	\\
$T$             & Temperature                                          	& K                 	\\
$v$             & Gas velocity along pipe                            	& m/s               	\\
$u$				& Radial mean velocity toward wall                     	& m/s               	\\
$u_T$		    & Gas molecule thermal velocity	  					   	& m/s		       	   	\\
$x$             & Coordinate along the tube                            	& m                 	\\
                &                                                       &                       \\
$Greeks$        &                                                      	&                   	\\
$\delta$        & Thickness of the SN$_2$ layer                        	& m                 	\\
$\varepsilon$   & Specific internal energy                             	& J/kg              	\\
$\rho$          & Density                                              	& kg/m$^3$          	\\
$\mu$           & Viscosity                                            	& Pa$\cdot $s       	\\
$\sigma_c$	    & Condensation coefficient							   	&			        	\\
$\sigma_e$	    & Evaporation coefficient							   	&					    \\
$\Gamma$		& Velocity distribution correction factor           	&					    \\
&                                                      				  	&                  	    \\
$Subscripts$    &                                                     	&                   	\\
$g$             & Bulk gas                                            	&                   	\\
$s$             & Surface of the frost layer                           	&                   	\\
$w$             & Copper tube wall                                    	&                   	\\
$SN$         	& Solid nitrogen                                      	&                   	\\
\hline
\hline
\end{xtabular}

\section*{References}
\bibliographystyle{elsarticle-num-names}
\biboptions{sort&compress}
\bibliography{References-pt3}

\end{document}